\documentclass[twocolumn,showpacs,aps,pre]{revtex4}
\usepackage{graphicx}
\usepackage{dcolumn}
\usepackage{bm}

\begin{document}

\title{Theory of De-Pinning of Monolayer Films Adsorbed on a Quartz Crystal Microbalance} 
\author{J. B. Sokoloff$^*$ and I. Webman$^{**+}$}
\address{$^*$Physics Department and Center for Interdisciplinary Research on
Complex Systems, Northeastern University, Boston, MA 02115; $^{**}$Physics 
Department, Bar-Ilan University, Ramat-Gan, Israel, $^+$Deceased.}
\date{\today }

\begin{abstract}

In quartz crystal microbalance (QCM) studies of the friction between an adsorbed monolayer film 
and a metallic substrate, the films are observed 
to slide relative to the substrate under inertial forces of order $10^{-14}dyn$ per film atom, 
a force much smaller than all theoretical estimates of the force that surface 
defects are capable of exerting. In this letter we propose, in order to 
resolve this issue, that if the defect potentials have a range of 
greater than an atomic spacing, the net force on a relatively stiff film due to the defects 
is likely to be extremely small. Line defects (e.g., step and facet edges and grain boundaries) 
as well as more localized defects (e.g., vacancies) are considered.    
\end{abstract}


\maketitle

Quartz crystal microbalance (QCM) studies of mono-layers of molecules 
on metallic substrates[1] provide detailed information about friction 
at the atomic level. The QCM consists of a  quartz oscillator of frequency 
$\sim 10^6 Hz $. Monolayer films adsorbed on the metallic surface of the quartz 
crystal oscillator are found to slip with respect to the surface during a period 
of the oscillator. Small changes observed in the frequency and in the associated 
Q factor allow us to gauge the 
friction between mono-layer  and substrate.  The amount of 
dissipation generated under most conditions implies that sliding motion of more 
than a lattice constant occurs during a period of the QCM. 

 For film atom mass $m\approx 10^{-22}g$, 
appropriate for xenon atoms, QCM frequency $\omega\approx 10^7 rad/s$ and 
amplitude $A\approx 100 A^o$, which are appropriate parameters for the QCM 
experiment, the inertial force $m\omega^2 A$ is only about $10^{-14}$ dyn 
per film atom. Calculations 
of pinning of a monolayer film by defects, based on both perturbation 
theory[2] and on molecular dynamic simulations[3]  
imply that this inertial force per atom due to oscillations of the substrate 
should not be sufficiently strong to overcome the pinning 
resulting from the defects. This is in fact what was found in 
experimental studies by Taborek[4] to attempt to reproduce 
the striking temperature dependence of the sliding friction found by 
Dayo and Krim for a nitrogen film sliding on a lead substrate, as 
the temperature dropped below the superconducting transition temperature 
of the lead[5]. It has been established [6], however, that the substrates 
used in this experiment were much more contaminated than those used in Ref. 5. 
In the usual treatment of an elastic 
medium interacting with a disordered potential \cite{larkin}, the range 
of the defect potential is assumed to be sufficiently small compared to 
a film lattice spacing, so that it can only interact with one film atom at a time. 
In contrast, pinning defects on the substrate typically extend  
over one or more atomic lengths. For example, the potential produced by 
line defects (e.g., step and facet edges and grain boundaries) clearly extend 
a number of lattice spacings along the length of the defect, and even a 
more localized defect such as 
a vacancy extends at least out past its nearest neighbor atoms, because 
the neighboring atoms will displace towards the missing atom, resulting 
in a potential which extends out at least that far. It will be 
argued here that for potentials, typical of surface defects, that 
have a long enough range to interact with two or more atoms along the 
sliding direction, the forces on 
these atoms tend to cancel each other, reducing the force on the film due 
to a single defect to 
a value much smaller than the maximum force that a defect can exert 
on a single film atom. 

It has been established using simulations \cite{tomassone} that the Larkin 
length\cite{larkin}, which is a measure of the stiffness of the film, 
is many lattice spacings long implying that the film distorts over 
distances long compared to the mean defect potential spacing, and hence 
certainly long compared to the defect potential width. It will now be 
demonstrated that because of the film stiffness, even for defect potentials 
with a range of a little more than a film lattice sp acing, the forces 
exerted on the film atoms tend to cancel to a great degree. 
(Incidentally, the simulations reported in Ref. 3 used a defect potential 
of much shorter range.)  Consider the potential energy of a rigid film of 
atoms interacting with a single defect potential 
$$\sum_{\bf R} v({\bf R}+\Delta {\bf r}) \eqno (1)$$ 
where $v({\bf R}+\Delta {\bf r})$ is the potential energy of an atom 
located at the point ${\bf R}+\Delta {\bf r}$ in the film due to the defect, 
${\bf \Delta r}$ represents a displacement of the film relative to the                  
potential and {\bf R} is an atomic position in the periodic lattice 
of the film. Since this quantity is a periodic function of $\Delta {\bf r}$ 
with the periodicity of the film lattice, and hence, it can 
be expressed as a Fourier series\cite {ashcroft} in the coordinates 
in the plane parallel to the interface  
$$\sum_{\bf R} v({\bf R}+\Delta {\bf r})=
\sum_{\bf G} \bar{v} ({\bf G},z)e^{i{\bf G}\cdot\Delta {\bf r}}
\eqno (2)$$
where the Fourier coefficient at a distance z 
from the surface, $\bar{v} ({\bf G},z)=
\Omega^{-1}\int d^2 r e^{-i{\bf G}\cdot {\bf r}}v({\bf r})$, where $\Omega$ 
is the unit cell area and {\bf G} is a reciprocal lattice vector of the 
film. The integral is taken over the whole film. For line defects, such as 
step and facet edges and grain boundaries v({\bf r}) will vary quite slowly as 
a function of {\bf r} along the length of the line defect (defined as the distance 
over which the defect is relatively straight). As a consequence, 
we shall see that the Fourier coefficient $\bar{v} ({\bf G},z)$ are likely to 
be quite small for general directions of {\bf G}, such that {\bf G} has a 
component along the length of the defect sufficiently large compared with 
$2\pi$ divided by the length of the defect, a condition which is easy to 
satisfy for any line defect long compared to a film lattice constant.  
The Fourier coefficients will also be very small for defects such as 
vacancies which are localized around a point on the substrate. 

Since the exact form of the defect potential is not known, let us 
illustrate this effect by studying  
a couple of simple smooth potentials which drop off 
reasonably rapidly at large distances, in order to develop a picture 
of what one expects for defect potentials with a range greater than 
a lattice spacing. Here we will consider primarily a two dimensional Gaussian 
and  Lorentzian potential, which represent two extremes, as the Gaussian falls to 
zero very rapidly at 
large distances while the Lorentzian falls off relatively slowly (as is evidenced by the 
fact that its second moment is infinite). For an anisotropic Gaussian defect potential, 
$v({\bf r})=-V_0 (z) e^{-[(x/b_1)^2+(y/b_2)^2]},$ where $b_1$ and $b_2$ are 
the range parameters, we obtain, for a triangular lattice 
$\bar{v} ({\bf G})= -2V_0 \pi (b_1 b_2/(3)^{1/2} a^2)e^{-(G_x^2 b_1^2/4+G_y^2 b_2^2/4)}$
For the circularly symmetric two dimensional Lorentian potential $v({\bf r})=-V_0 (z) [1+(x/b_1)^2+(y/b_2)^2]^{-1}$, $\bar{v} ({\bf G},z)$ 
on the right hand side of Eq. (2) is equal to $-[4\pi b_1 b_2/a^2 (3)^{1/2}]V_0 K_0 (Q)$, 
where $Q=(G_x ^2 b_1^2+G_y^2 b_2^2)^{1/2}$ and $K_0 (Q)$ is the spherical Hankel function 
of the first kind with an imaginary argument [9], whose large argument assymptotic form is 
$K_0 (Q)\approx (2/\pi Q)^{1/2} e^{-Q}$. 
These forms can be used to model a line defect which runs along the x-axis if 
we take $b_1>>b_2$, with $b_1$ representing the distance along the defect length that 
one must travel in order to reach a point at which the relatively straight section of the 
defect ends, and $b_2$ 
represents the width of the defect potential perpendicular to the direction of the line 
defect. We will assume the film's crystallographic axes and sliding direction 
to be at an arbitrary angle with respect to the x and y axes. If we wish to model 
localized defects, such as vacancies or interstitials, we set $b_1=b_2$. The magnitude of the 
Fourier coefficient is a very sensitive function of the range of the potential. 
By taking the negative of the gradient of Eq. (2) with respect to $\Delta {\bf r}$, we obtain the force on the film, whose maximum value is approximately $G\bar{v} ({\bf G},z)$, where {\bf G} 
is one of the smallest reciprocal lattice vectors. The maximum force on the film 
for the Gaussian defect potential for $b_1=b_2=b=1.23a$ is $0.979\times 10^{-7}V_0/b$ and for 
$b_1=b_2=b=0.5a$ it is $0.0248V_0/b$. The corresponding value for the maximum force for the 
two dimensional Lorentzian is $0.00349V_0/b$ for $b_1=b_2=b=1.23a$ and $0.0729 V_0/b$ 
for $b_1=b_2=b=0.5a$. These quantities should be compared to the maximum possible force 
that these model potentials can exert on a single film atom, which is $0.429V_0/b$ for 
the Gaussian and $0.5 V_0/b$ for the Lorentzian. We see that for both of these values of b 
the maximum force on the film is much smaller than the maximum force that it can exert on a 
single atom for both model potentials and it is a very sensitive function of b/a. 

For line defects, the effects can be even more dramatic. For example, for a line defect 
along the x-axis, it is reasonable to assume that $b_1$ is a number of lattice spacings long, 
and hence, considerably longer than a film lattice spacing. 
Hence, for virtually all orientations of the film axes, $|G_x b_1|>>1$, and hence, 
we can see for the two dimensional Gaussian and Lorentzian potentials discussed in 
the last paragraph, their Fourier coefficients will be extremely shall, and hence we see from 
Eq. (2), the dependence of the interaction of the potential with the film with the defect 
on $\Delta {\bf r}$ will be extremely small, implying extremely small forces on the film.  

Other model defect potentials give similar results. For example, the potentials 
$-V_0/[cosh (x/b_1)cosh(y/b_2)]$ and $-V_0/[(1+(x/b_1)^{2n} (1+(y/b_2)^{2n}]$, 
where n is any integer, also  result 
in Fourier coefficients which fall off as exponential functions for reasonably large values 
of $|G_x| b_1$ and $|G_y| b_2$. The latter potential with a sufficiently large value of n 
is a good model for a line defect whose potential is nearly constant along most of its length 
and only has significant variation near its ends. The only requirement in order 
to get a Fourier coefficients which fall off rapidly at large values of its argument (i.e., exponentially or better) is that the potential due to the defect be smooth.

%In order to treat a lattice which is not completely stiff, we replace n in 
%Eq. (1) by $n+u (n+x),$ where $u (n+x)$ is the displacement due to distortion 
%induced by the defect potential of the $n^{th}$ atom. Since we are considering a 
%stiff lattice, we can expand v to first order in u. Then the right hand sides of 
%Eqs. (3) and (4) become
%$$\sum^{\infty}_{s=-\infty} (-1)^s \int^{\infty}_{-\infty} dy [v(y)+$$
%$${dv\over dy} u(y)]e^{-i2\pi sy} e^{-2\pi si (x-1/2)}. \eqno (5)$$
%Since both v(y) and ${dv(y)\over dy}$ are both functions that fall to zero 
%when y becomes large compared to b, we expect both therms in the Fourier 
%transform (i.e., the integral over y) to become small as the product sb 
%increases. Hence our conclusion for a rigid lattice that the x dependent 
%term in the total interaction of the chain with the defect potential is 
%much smaller than $V_0$ for values of b which are not much larger than 1 
%should still hold. 

At any temperature, there will be lattice vibrations. At a given 
instant of time, a lattice vibration of wavevector {\bf q} will add a term 
${\bf A} cos({\bf q}\cdot ({\bf r}+{\bf R})-\omega t)$ to {\bf R} in 
Eqs. (1), where {\bf A} is the amplitude (assumed to be much smaller than 
a lattice constant a) and $\omega$ is the frequency of the vibrational mode 
and t is the time. Then, expanding v in a Fourier series, Eq. (1) becomes 
$$\sum_{\bf r} N^{-1}\sum_{\bf k} \bar{v} ({\bf k},z) e^{i{\bf k}\cdot 
({\bf R}+\Delta {\bf r})} e^{i{\bf k}\cdot {\bf A} 
cos({\bf q}\cdot ({\bf r}+{\bf R})-\omega t)}, \eqno (3)$$
where N is the number of atoms in the film. 
Expanding the second exponential in Bessel functions Eq. (3) becomes 
$$N^{-1}\sum_{\bf R}\sum_{\bf k} \bar{v} ({\bf k},z)\sum_{n=0}^{\infty} i^n J_n ({\bf k}\cdot {\bf A}) 
e^{i({\bf k}-n{\bf q})\cdot ({\bf R}+\Delta {\bf r})} e^{i n\omega t}, 
\eqno (4)$$
which when summed over {\bf R} becomes
$$\sum_{\bf G}\sum_n i^n J_n ({(\bf G}-n {\bf q})\cdot {\bf A})
\bar{v} ({\bf G}-n{\bf q},z) e^{i({\bf G} \cdot\Delta {\bf r}-n\omega t)}. \eqno (5)$$
For the n=1 term, which should be a good approximation for the small values 
of $|({\bf G}-n{\bf q})\cdot {\bf A}|$ characteristic of 
lattice vibrations, we find that $\bar{v} ({\bf G}-{\bf q},z)$ for 
the Gaussian potential now contains a term $e^{-|{\bf G}-{\bf q}|^2 b^2/2}$, 
which can be considerably larger than  $e^{-G^2 b^2/2}$, which occurs for n=0, by a sufficient amount 
to make the former term dominate over the latter for values of q 
which are comparable in magnitude to the smallest values of G, even 
though it is multiplied by $J_n ({(\bf G}-{\bf q})\cdot {\bf A})$, 
which is much less than 1. As the former term is proportional to 
{\bf A} which oscillates in time for a lattice vibration, this term 
produces an oscillating term in the 
force on the film, which can be 
larger in magnitude than the force on the film 
that would occur if there were no vibrations 
[i.e., the n=0 term in Eq. (5)]. Thus at nonzero  
temperatures, the lattice vibrations can produce oscillating 
forces acting on the part of the film in the range of the potential, 
which dominate over the static pinning force 
on the film due to the potential under consideration, 
implying that at nonzero temperatures, the film will 
not be pinned by the defects because the film is pushed out of its 
total potential minimum by this force. 

In the previous paragraphs, the film was taken to be completely rigid, and 
the distortion of the film resulting from its interaction with the 
defect potential was neglected. In order to examine the validity of this 
approximation, let us now consider the small distortion of the film 
resulting from the potential, in order to determine whether it will destroy 
the reduction of the potential's interaction with the film discussed in the 
previous two paragraphs. The displacement of a film atom located at 
point {\bf R} is given in lowest order perturbation theory by
$${\bf u} ({\bf R})\approx-\sum_{\bf R'} {\bf G} ({\bf R}-{\bf R'})
\cdot\nabla' v({\bf R'},z),\eqno (6)$$
where ${\bf G} ({\bf R}-{\bf R'})$ is the elasticity Green's function 
\cite {ashcroft}. It is shown in Ref. \cite {ashcroft} that 
$${\bf G} ({\bf R}-{\bf R'})=N^{-1}\sum_{{\bf k},\alpha}
{\hat{\epsilon}_{{\bf k},\alpha}\hat{\epsilon}_{{\bf k},\alpha}
e^{i{\bf k}\cdot ({\bf R}-{\bf R'})}\over \omega^2_{\alpha} ({\bf k})},
\eqno (7)$$
where $\omega_{\alpha} ({\bf k})$ is the frequency of the $\alpha^{th}$ 
vibrational mode of wavevector {\bf k} and $\hat{\epsilon}_{{\bf k},\alpha}$ 
is the unit vector which gives its polarization. Substituting Eq. (6) 
in Eq. (6), we obtain 
$${\bf u} ({\bf R})\approx -N^{-1} i\sum_{{\bf k},\alpha}
{\hat{\epsilon}_{{\bf k},\alpha}\hat{\epsilon}_{{\bf k},\alpha}
e^{i{\bf k}\cdot {\bf R}}\over \omega^2_{\alpha} ({\bf k})}\cdot 
{\bf k} \bar{v} ({\bf k},z). \eqno (8)$$
If we add {\bf u} to {\bf R} in Eq. (1), linearize in {\bf u} and 
substitute for {\bf u} from Eq. (9), we find that the first order term in 
an expansion in powers of {\bf u} for the force on the film is given by
$$N^{-1}\sum_{{\bf k},{\bf G}}{[({\bf G}-{\bf k})\cdot\hat{\epsilon}_{{\bf k},\alpha}]
[{\bf k}\cdot\hat{\epsilon}_{{\bf k},\alpha}]\over\omega^2_{\alpha} ({\bf k}})
\bar{v} ({\bf G}-{\bf k})\bar{v} ({\bf k})e^{i{\bf k}\cdot\Delta {\bf r}}.
\eqno (9)$$ 
We see from Eq. (9) that since we found earlier that  
$\bar{v} ({\bf k},z)$ becomes 
negligibly small rapidly once k becomes larger than 1/b, 
only Fourier components of wavevector larger than 1/b 
will make a significant contribution to Eq. (9). 
We shall now illustrate that the modulation of the film resulting from the 
defect potential should have little effect on our conclusion that the defect interacts 
extremely weakly with the adsorbed film with a simple calculation. 
Consider a stiff two dimensional 
square lattice of atoms of lattice constant a interacting with a Gaussian 
potential $-V_0 e^{-r^2/b^2}$, were r is the distance from the center of 
the potential. We choose the range parameter of the potential, 
b=1.23a. For this choice of parameters the potential range is 
about one and a half lattice spacings. The lattice 
is slid a distance s and the total force due to this potential acting on 
it is plotted as a function of s (the lower amplitude curve in Fig. 1). 
The position of an atom in the lattice is then displaced by an amount 
$0.01acos((2\pi/\lambda)x_n)$ in the x-direction,  where the position 
of a film atom is $(x_n,y_m)=(na,ma)$ where n and m are integers,   
to simulate the effect of this static modulation of the film. There is no displacement $\Delta x$ included because a modulation 
produced by the potential does not slide relative to the potential. Here, $\lambda$ is 
chosen to be 4b (in the range of the shortest $\lambda$ value which makes a significant contribution 
in Eq. (10)), so that half of a wavelength extends across the width of  the 
potential well. The results are given in Fig. 1.  
As can be seen the force on the film when such a modulation is present is comparable 
to that in the absense of a modulation, and is much less than the maximum force that a 
defect can exert on a film atom.

\begin{figure}[tbp]
\center{
\includegraphics [angle=0,width=3.4in]{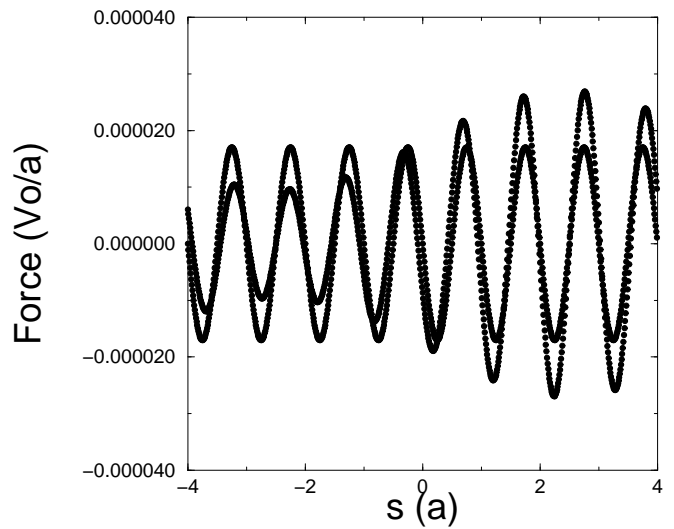}}
 \vskip -0.3cm
%\centereps{2.05in}{2.05in}{lubricated-surface3.eps} \vskip -0.3cm
\caption{The total force due to a Gaussian potential acting on a rigid 
lattice  of atoms of spacing a is plotted as a function of the displacement 
of the film s in the lower amplitude curve for b=1.236067978a. The total 
force acting on a modulated lattice is shown in the higher amplitude curve.} 
\label{fig1}
\end{figure}

We have considered the interaction of a single defect with the film and have 
demonstrated that this can easily be much weaker than the maximum possible 
interaction of that defect with a single film atom. When we consider the 
interaction of the film with the random distribution of defects over the 
substrate that is typically found, we must consider the fact that in the 
weak pinning regime (in which the film is not able to distort enough to 
minimize its interaction with all of  the randomly distributed 
defects), the forces due to the defects 
do not act in phase \cite {larkin}. As a consequence, the net force per atom on the 
film gets reduced in addition by a factor of the square root of 
the number of defects within a Larkin domain, which is proportional to the ratio of the Larkin 
length and a film lattice constant. 
Since for a two dimensional solid  the Larkin length is equal to the ratio of the 
elastic constant and the interaction with a single defect\cite {lee}, the over-all interaction 
of the defects with the film is proportional  to the square of the defect interaction, rather 
than being a linear function of it. Furthermore, our single defect interaction estimates 
are in fact an upper bound on this quantity.

In conclusion, we have proposed a mechanism which may explain why local defects, 
which must certainly be present on even the smoothest surfaces, might not 
prevent a stiff monolayer film from sliding under the extremely weak inertial 
forces that occur in a quartz microbalance experiment, if 
the defect potential has a range greater than about 1.5 film lattice spacings. 
Our proposed mechanism is supported by a simple model calculation, based on the fact 
that each defect can interact with several film atoms at a time. As a consequence, 
the resultant forces on the film are much less than the maximum force provided by the 
defect potential. Lattice vibrations are able 
to provide an oscillating force which is larger than the net 
static force due to 
the defect potential acting on the film. This can lead to a high 
degree of thermal activation of the 
sliding of the film.  
\section*{Acknowledgments}
J.B. Sokoloff wishes to thank the Department of Energy 
(Grant DE-FG02-96ER45585).

\end{document}